\documentclass[10pt,letterpaper]{article}
\usepackage[top=0.85in,left=2.75in,footskip=0.75in]{geometry}

\usepackage{amsmath,amssymb}

\usepackage{changepage}

\usepackage{textcomp,marvosym}

\usepackage{cite}

\usepackage{nameref,hyperref}

\usepackage[nopatch=eqnum]{microtype}
\DisableLigatures[f]{encoding = *, family = * }

\usepackage[table]{xcolor}

\usepackage{array}

\newcolumntype{+}{!{\vrule width 2pt}}

\newlength\savedwidth

\raggedright
\usepackage[aboveskip=1pt,labelfont=bf,labelsep=period,justification=raggedright,singlelinecheck=off]{caption}

\makeatletter
\renewcommand{\@biblabel}[1]{\quad#1.}
\makeatother

\usepackage{lastpage,fancyhdr,graphicx}
\usepackage{epstopdf}
\fancyheadoffset[L]{2.25in}
\fancyfootoffset[L]{2.25in}
\newcommand{\orcid}[1]{\href{https://orcid.org/#1}{#1}}

\begin{document}
\vspace*{0.2in}

\begin{flushleft}
{\Large
\textbf\newline{10 simple rules for making your software outlive your job}
}
\newline
\\
Richard Littauer\textsuperscript{1*},
Greg Wilson\textsuperscript{2},
Jan Ainali\textsuperscript{3},
Eman Abdullah AlOmar\textsuperscript{4},
Sylwester Arabas\textsuperscript{5},
Yanina Bellini Saibene\textsuperscript{6},
Kris Bubendorfer\textsuperscript{7},
Kaylea Champion\textsuperscript{8},
Clare Dillon\textsuperscript{9},
Jouni Helske\textsuperscript{10},
Pieter Huybrechts\textsuperscript{11},
Daniel S.\ Katz\textsuperscript{12},
Chang Liao\textsuperscript{13},
David Lippert\textsuperscript{14},
Fang Liu\textsuperscript{15},
Pierre Marshall\textsuperscript{16},
Daniel R.\ McCloy\textsuperscript{17},
Ian McInerney\textsuperscript{18},
Mohamed Wiem Mkaouer\textsuperscript{19},
Priyanka Ojha\textsuperscript{20},
Christoph Treude\textsuperscript{21},
Ethan P.\ White\textsuperscript{22}
\\
\bigskip

\textbf{1} CURIOSS, SustainOSS, GNOME Foundation, and Te Herenga Waka Victoria University of Wellington, Te Whanganui-a-Tara Wellington, Aotearoa New Zealand, \orcid{0000-0001-5428-7535}\\
\textbf{2} Third Bit, Toronto, ON, Canada, \orcid{0000-0001-8659-8979}\\
\textbf{3} Open By Default, \orcid{0000-0001-8747-1670}\\
\textbf{4} Department of Systems and Enterprises, Stevens Institute of Technology, Hoboken, NJ, USA, \orcid{0000-0003-1800-9268}\\
\textbf{5} AGH University of Krakow, Poland, \orcid{0000-0003-2361-0082}\\
\textbf{6} rOpenSci, Santa Rosa, LP, Argentina, \orcid{0000-0002-4522-7466}\\
\textbf{7} Te Herenga Waka Victoria University of Wellington, Wellington, New Zealand, \orcid{0000-0003-4315-8337}\\
\textbf{8} Computing and Software Systems, University of Washington Bothell, Bothell, WA, USA, \orcid{0000-0001-6196-942X}\\
\textbf{9} CURIOSS, Lero, and University of Galway, Ireland, \orcid{0009-0008-6205-0296}\\
\textbf{10} INVEST Research Flagship Centre, University of Turku, Finland, \orcid{0000-0001-7130-793X}\\
\textbf{11} Research Institute for Nature and Forest (INBO), Brussels, Belgium, \orcid{0000-0002-6658-6062}\\
\textbf{12} University of Illinois Urbana-Champaign, IL, USA, \orcid{0000-0001-5934-7525}\\
\textbf{13} Atmospheric, Climate, and Earth Sciences, Pacific Northwest National Laboratory, Richland, WA, USA, \orcid{0000-0002-7348-8858}\\
\textbf{14} Open Source Program Office, George Washington University, Washington, DC, USA, \orcid{0009-0003-6444-9595}\\
\textbf{15} Georgia Institute of Technology, GA, USA \orcid{0000-0002-3383-2191} \\
\textbf{16} Oxford University, UK, \orcid{0000-0001-9245-7670}\\
\textbf{17} Institute for Learning \& Brain Sciences, University of Washington, Seattle, WA, USA, \orcid{0000-0002-7572-3241}\\
\textbf{18} Department of Mechanical Engineering, Imperial College London, London, UK, \orcid{0000-0003-2616-9771}\\
\textbf{19} Department of Computer Science, University of Michigan-Flint, Flint, MI, USA, \orcid{0000-0001-6010-7561}\\
\textbf{20} \orcid{0000-0002-6844-6493}\\
\textbf{21} Singapore Management University, Singapore, \orcid{0000-0002-6919-2149}\\
\textbf{22} Department of Wildlife Ecology and Conservation, University of Florida, Gainesville, FL, USA, \orcid{0000-0001-6728-7745}\\
* richard@burntfen.com
\end{flushleft}

\noindent
We are grateful for contributions from
Tobias Augspurger,
Bill Branan (\orcid{0000-0002-4735-6624}),
Paola Corrales (\orcid{0000-0003-1923-912}9),
Sam Cunliffe (\orcid{0000-0003-0167-8641}),
David Eyers (\orcid{0000-0002-7284-8006}),
Elena Findley-de Regt,
Tommy Guy (\orcid{0009-0003-3652-5036}),
Jonathan Guyer (\orcid{0000-0002-1407-6589}),
Mala Kumar (\orcid{0009-0004-3619-1037}),
Geoffrey Lentner (\orcid{0000-0001-9314-0683}),
Georg Link (\orcid{0000-0001-6769-7867}),
Daniel Morillo-Cuadrado (\orcid{0000-0003-3021-3878}),
David Pérez-Suárez (\orcid{0000-0003-0784-6909}),
Phani Velicheti (\orcid{0009-0004-2580-3624}),
and everyone else who provided feedback along the way.

\section*{Abstract}

Loss of key personnel has always been a risk for research software projects.
Key members of the team may have to step away due to illness or burnout,
to care for a family member,
from a loss of financial support,
or because their career is going in a new direction.
Today,
though,
political and financial changes are putting large numbers of researchers out of work simultaneously,
potentially leaving large amounts of research software abandoned.
This article presents ten rules to help researchers ensure that
the software they have built will continue to be usable
after they have left their present job---whether in the course of voluntary career moves or researcher mobility,
but particularly in cases of involuntary departure due to political or institutional changes.


\section*{Introduction}

Academic life is often uncertain,
particularly for those who work on research software,
an asset that is essential to modern research \cite{Pearson2025}
but often underfunded, undercited, and overlooked \cite{Carver2022}.
The situation has recently worsened dramatically due to changes in the political and economic climate:
researchers in all disciplines face the loss of funding or jobs,
a smaller pool of incoming students,
and institutional policies that prioritize commercial returns.
Researchers in non-academic organizations
such as government laboratories, non-governmental organizations (NGOs), and non-profits are also affected \cite{Woodward2025}
as national research strategies are changed
and associated public funding is redirected or cut entirely \cite{Nature2019,Nature2021,Nature2023,RatRisk2024,Nature2025}.

Historically,
open-source research software has been longer-lived than other kinds of open-source software 
and government involvement has been associated with longer lifespan \cite{Thakur2025}. 
The challenges of responding to macro-level changes in software development support may be therefore new and unfamiliar to research software developers.
In this changing environment,
research software projects must be \emph{resilient}.
Code should outlast situations such as the author changing workplaces or careers,
partnerships failing,
a lab closing down,
or tooling and digital infrastructure becoming unavailable.
We therefore present ten rules to ensure that your software remains usable by others
even if you are no longer able to work on it.
Some of these rules are most useful when adopted at the beginning of a project
to build resilience over time (e.g., Rules 1, 2, 5, and 7),
while others are particularly relevant when a transition or disruption is imminent
(e.g., Rules 3 and 8).
You do not need to adopt all of them to make things better;
instead,
as in all emergencies,
you should do what you can,
where you are,
with what you have.

This guide focuses on software.
Other guides already exist for data \cite{Perkel2023},
where archives such as \href{https://zenodo.org/}{Zenodo},
\href{https://figshare.com/}{FigShare},
the \href{https://eotarchive.org/}{End Of Term Archive}
or the \href{http://archive.org/}{Internet Archive}
are now widely used to ensure long-term availability.
However,
we have not found any guide to make research software resilient
or to \emph{sunsetting} it (i.e., phasing it out gracefully).
Even if your job is not in danger,
these rules will help others access and use your code,
contribute to your project,
and to reproduce and cite your work.
As such,
our recommendations
are aligned with the global push for a more equitable assessment of research contributions
by initiatives such as \href{https://sfdora.org/}{DORA},
\href{https://coara.eu/}{coARA},
and \href{https://adore.software/}{ADORE.software}.

\begin{quote}
  \noindent
  \textbf{Collecting wisdom}

  In 2013,
  the government of Canada ordered the closure of
  five of the Department of Fisheries and Oceans' seven libraries \cite{Nikiforuk2013}.
  No one knows how many thousands of volumes of irreplaceable environmental records were destroyed,
  but even scientists who were directly affected did not make plans in case something similar happened again.
  In contrast,
  when faced with the prospect of a populist demagogue becoming president,
  researchers in Argentina knew from their collective experience what might happen next
  and took action accordingly \cite{DelBoca2024}.

  We left many suggestions out of this paper due to space constraints
  or because we felt they would age poorly.
  What worries us is the ones we didn't include because
  we have been lucky enough not to have to think about them.
  In an effort to correct that,
  we have set up a task force hosted by \href{https://curioss.org}{CURIOSS},
  \href{https://sustainoss.org}{SustainOSS},
  and the \href{https://www.researchsoft.org//}{Research Software Alliance},
  and plan to run monthly meetings
  to share ideas and support one another.
  If you have suggestions,
  please contribute them at \url{https://www.researchsoft.org/tf-succession-planning/}.
\end{quote}

\section*{Rule 1: Stay within the law}

Ensuring your work remains usable is not worth putting yourself at legal risk.
Before following any of the rules below,
make sure you have a legal right to do so.
Most institutions and journals now have policies for licensing code and/or releasing it publicly \cite{Katz2018,Ham2019}.
Funding agencies at various levels also often have policies
(e.g.,
EU\footnote{\url{https://web.archive.org/web/20250424181252/https://commission.europa.eu/about/departments-and-executive-agencies/digital-services/open-source-software-strategy_en} (archived 2025-04-24)},
UKRI\footnote{\url{https://web.archive.org/web/20250422025105/https://www.ukri.org/manage-your-award/publishing-your-research-findings/making-your-research-data-open/} (archived 2025-04-22)},
and NASA\footnote{\url{https://web.archive.org/web/20250422005753/https://science.nasa.gov/open-science/nasa-open-science-funding-opportunities/} (archived 2025-04-22)})
which may or may not align with those of institutions and journals,
and institutional policies may place restrictions on what you are allowed to say publicly about your situation.
Our first actionable rule is therefore to find out what those policies are,
e.g.,
whether you need explicit permission to publish your software,
and if so,
from whom.

If you are unsure whether there is a policy or not,
ask your direct manager,
the person who pays you,
or offices like the research office,
the library,
or the technology transfer office.
If your institution has a dedicated
\href{https://sustainoss.org/academic-map/universities/index.html}{Open Source Program Office} (OSPO),
ask them
or connect to networks like \href{https://curioss.org}{CURIOSS},
the \href{https://ospo-alliance.org/}{OSPO Alliance},
or the \href{https://todogroup.org/}{TODO Group}
that advocate for OSPOs.
If you work with multiple institutions,
look at your contract and the bilateral agreements around products or deliverables.

Once you know whether your institution has a policy,
get whatever sign-offs you need immediately.
Reach out to colleagues who can review your code if that is necessary for publishing it,
and review theirs if asked in return.
If it appears that there is no formal sign-off process,
send an email to someone in authority
(e.g., the chair of your department or your grant officer)
saying that you believe this to be the case,
and copy that email to an account you will be able to access
after leaving your institution.

Do not assume that if you had permission before,
you have it now or will have it in the future.
Policies are changing rapidly,
and you may find yourself locked into one that no longer allows you to do what you want.
Also consider that the person you report to may be replaced with one who knows you less well or is less sympathetic,
so acting now may be easier than acting later.

Finally,
if you have your next job lined up,
ask about their policies
and make sure that your right to share your work is written into your contract.
Making your code comply with policy after the fact is riskier and more time-consuming than doing so early,
and sometimes not possible.

\begin{quote}
  \noindent
  \textbf{Not all rules are created equal}

  Anyone who has worked in a large bureaucratic organization knows
  the difference between the rules as written and the rules as enforced.
  If you are leaving your position on short notice or under difficult circumstances,
  you may want to be selective about which forms you fill in proactively
  and which you just haven't gotten around to yet.
  Members of marginalized groups have often had more practice with these tactics
  than members of privileged groups \cite{Scott1987};
  if you belong to the latter,
  a quiet conversation or two with selected colleagues may help you see your priorities more clearly.
\end{quote}

\section*{Rule 2: Define your threat model}

When making plans, it's useful to know what you're planning \emph{for}.
A \textbf{threat model} is a structured analysis of where threats might come from,
what form they might take,
what social and technical vulnerabilities they might exploit,
and how they might be mitigated or resolved \cite{Torr2005}.
Being explicit about threat models helps you prioritize,
build consensus with colleagues,
and check whether you have forgotten something important.
As with all disaster planning,
making plans before you need them may help you identify risks you can eliminate proactively.

\begin{enumerate}
\item
  \textbf{Individual threats} affect one or a few members of your team,
  such as an international student having their visa revoked without notice, or a contributor taking extended leave.
  Especially for software under the maintenance of a single author,
  individual threats can also come from more mundane career or life changes.
  The most common way to prepare for this is to require everyone to document their work,
  but that rarely works in practice:
  \begin{enumerate}
  \item
    The hours spent writing those descriptions are hours \emph{not} spent doing research,
    so people will always short-change the former to focus on the latter.
  \item
    People invariably fail to write down the ``obvious'' parts of their work
    that are anything but obvious to the next person.
  \end{enumerate}
  Rule~8 explores alternatives,
  particularly ones that can be put in place on short notice.

\item
  \textbf{Leadership threats} are individual threats that affect the project's leader,
  such as the leader being doxxed or personally targeted in the media because of their work.
  Rule~3 discusses ways to plan for this.

\item
  \textbf{Institutional threats} affect large groups at once,
  such as your department being shut down
  or your entire field having funding cut.
  These events affect so many people at the same time
  that the rest of the community cannot absorb them.
  Regional and national governments plans for disasters like these
  by having evacuation plans to get victims to safe(r) places
  and corollary plans for putting beds in high school gyms
  and flying in food and emergency medical personnel to help people when they arrive.
  When government institutions themselves are in a period of upheaval, 
  mutual aid and civil society organizations may serve as an alternative.
  At the time of writing,
  universities and researchers' professional societies have not started to externally share equivalent planning, if they have started making those plans at all.

\item
  \textbf{Global threats} affect everyone,
  not just researchers.
  For example,
  there is no technical or legal obstacle for the US government
  requiring American companies to charge for video conferencing calls
  involving participants outside the United States.
  Similar levies on email,
  file storage,
  and other online services would undoubtedly prompt national governments to find alternatives,
  but would still result in (at least) months of disruption.

\end{enumerate}

\section*{Rule 3: Decide if you are ending, pausing, or handing off}

\begin{figure}[htbp]
  \centering
  \includegraphics[width=0.8\textwidth]{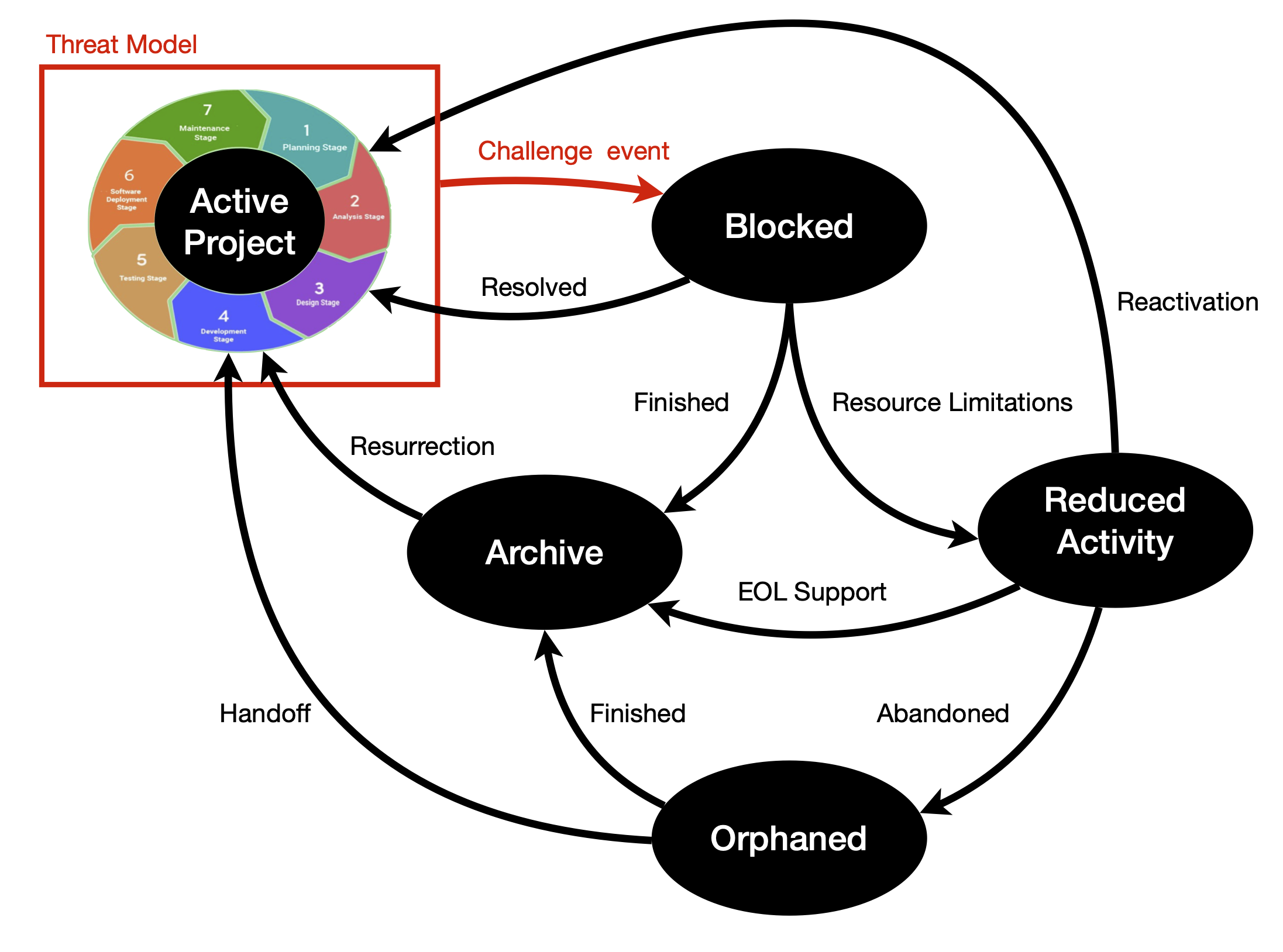}
  \caption{The software lifecycle.}
  \label{fig:lifecycle}
\end{figure}

Thousands of books describe how to start a business,
but only a handful discuss handing one over or winding one down,
and most of those focus on succession within family-owned businesses.
Guides to running a laboratory say little more \cite{Barker2010,Cohen2018},
so such wisdom as we have is passed on person-to-person if it is passed on at all. 
Consider the following augmented Software Lifecycle (SLDC) in Fig.~\ref{fig:lifecycle}, that is based on~\cite{deRSE25} and modified for this paper.

The first step is to decide if work on the project is ending,
being suspended temporarily (for some version of ``temporarily''),
or if you are handing it on to someone else.
If the project is ending or going on hiatus,
focus on making what you have accomplished findable and citable
as discussed in the rules that follow.
To help you prioritize,
write up the goals you initially had for this project
and ask how well they were served by what you actually did.
This approach could lead you to understand that the project does not have to conclude,
but can be integrated into someone else's work,
even without your continued participation.

If instead you hope to hand the project on to someone else,
you can either choose a successor
(who will typically be someone already working on the project)
or ask your peers for volunteers or recommendations.
Writing a \emph{succession plan} that describes what the work will actually entail
will help in both cases.
What will your successor have to learn and do during the transition?
What will they be responsible for once they are in charge?
And crucially,
how much work is running the project actually going to be?
Please be honest rather than optimistic:
finding out after you have agreed to take something on
that it is much larger than you realized
can strain both professional and personal relationships.

Some projects try to ensure longevity by working with a fiscal sponsor such as
\href{https://numfocus.org/}{NumFOCUS},
\href{https://sfconservancy.org/}{Software Freedom Conservancy},
\href{https://www.eclipse.org/}{the Eclipse Foundation},
\href{https://oscollective.org/}{Open Source Collective},
or other \href{https://sustainoss.org/academic-map/organizations/}{foundations and fiscal hosts}.
Through fiscal hosting,
a project can take donations from its community to fund continued maintenance and other costs.
Universities and governments are generally poorly suited for this due to high overheads and procurement costs;
fiscal sponsors may take an order of magnitude less.
In addition,
a fiscal sponsor places the software in a neutral home
where it is no longer owned by a single research organization,
which helps to insulate it from changes in policy.

However,
fiscal sponsorship is only an option in some countries,
and being accepted as a sponsored project is a long and challenging process.
Even when it is possible,
the copyright holder of the work may continue to be the institution that paid for the development of the software.
Private foundations may be a more attractive alternative in some jurisdictions;
as per Rule~1,
check with \emph{local} legal experts before taking action.

\section*{Rule 4: Choose an open license (if you can)}

Making code publicly available does not ensure that other people can use it;
you maintain all copyright unless you explicitly include a license,
so they may still need explicit permission
or may simply be unsure whether they do or not.
The Open Source Initiative (OSI)
maintains a list of \href{https://opensource.org/licenses}{open source licenses} it has approved;
the implications of these licenses are widely understood,
so choosing one of those will make your project more understandable to others.
Please do \emph{not} try to write your own license
or simply waive copyright and place the software into the public domain,
as what is meant by ``public domain'' varies from country to country.
(In countries such as France and Germany
it may not actually be possible to waive copyright and place a work in the public domain.)

Broadly speaking,
the MIT, Apache 2.0, or BSD-3-Clause licenses are \emph{permissive} licenses
that place the fewest restrictions on modification, distribution, and re-use.
The GPLv3 and AGPL-3.0 licenses are \emph{copyleft} licenses
that require people to share their modifications to the software with the community \cite{Morin2012}.
These copyleft licenses can prevent companies from taking advantage of work without giving anything in return,
but some institutions discourage or disallow their use out of fear that they will constrain commercialization,
and some companies have policies against using software with some copyleft licenses.

More broadly,
\href{https://creativecommons.org/}{Creative Commons} licenses can be used for documentation or research reports
but should be avoided for actual code,
while \href{https://ethicalsource.dev/}{ethical licenses}
or licenses based on the \href{https://blueoakcouncil.org/license/1.0.0}{Blue Oak Model License} may also be appropriate.
\href{http://choosealicense.com}{choosealicense.com} and other sources provide guidance \cite{Fogel2020,Fortunato2021};
discussion of why people \emph{don't} share is also helpful \cite{Gomes2022}.
Whatever license you choose,
place its text in a file called \texttt{LICENSE}, \texttt{LICENSE.txt}, or \texttt{LICENSE.md}
in the root directory of your project,
as this is where people and automated tools alike will look for it.

\begin{quote}
  \noindent
  \textbf{The law, again}

  Most discussion of open source licensing focuses on circumstances in a handful of affluent countries
  in the same way as discussion of fiscal sponsorship (Rule~3).
  For example,
  many Latin American countries have access to information laws that include research outcomes and artifacts.
  You can publish data and software under that,
  but the institution sets the license type,
  and you must use institutional repositories.
  As always,
  consult \emph{local} legal experts before taking action.
\end{quote}

\section*{Rule 5: Save everything in multiple places}

Now that you are legally able to share your code, remember LOCKSS: Lots of Copies Keep Stuff Safe.
But copies are not enough:
the threats to your project are political as well as technological (Rule~2),
so you should ensure that loss of institutional support
(or worse, that institution turning on you)
does not mean loss of access.

\begin{quote}
  \noindent
  \textbf{The road not taken}

  Fifteen years ago,
  a file-sharing system for scientific data based on the BitTorrent protocol
  failed to find wide adoption because of the (quite reasonable) association in many institutions' collective minds
  between BitTorrent and illegal downloading \cite{Langille2010}.
  At the time of writing,
  though,
  many individuals and groups are turning to BitTorrent to share datasets that are at risk of disappearing
  because of the protocol's resilience.
\end{quote}

\href{https://github.com/}{GitHub},
\href{https://gitlab.com/}{GitLab},
and \href{https://bitbucket.org/}{BitBucket}
are social coding platforms centered around Git,
a widely-used open source version control tool.
However,
they are all commercial entities based in the same legal jurisdiction,
which means they are vulnerable to \emph{correlated threats}:
trouble with any of them may mean trouble with all of them.
\href{https://codeberg.org/}{Codeberg},
while currently much smaller,
is a non-profit based in a different jurisdiction;
consider mirroring your work there or using it as your primary host.
Keep in mind that persistent code archival is not a feature of development platforms.
Repositories can be altered,
made private,
or deleted
(also as a result of government takedowns\footnote{\url{https://github.com/github/transparency/tree/main/data/government_takedowns}});
the platform itself may be shut down
(as happened to the non-profit Gna!\footnote{\url{https://en.wikipedia.org/wiki/Gna!}}),
or it may change its revenue model
(as happened with SourceForge\footnote{\url{https://en.wikipedia.org/wiki/SourceForge}},
which once dominated the open source repository space in the way that GitHub does today \cite{Tamburri2020}).

When you create projects on hosted services,
use a team account as the project owner rather than a personal account.
Doing this makes it easier to give other people permission to manage the project,
and as noted in Rule~6,
a project with multiple owners from different institutions
is harder for any one institution to lock down.
For the same reason,
do not rely solely on logins or email addresses associated with your institution:
instead,
ask yourself if you will still have access to the project if you lose that identity,
and add an identity you personally control to the team that owns the project.
You can also add your \href{https://orcid.org/}{ORCID} to the project's metadata
so that the project links to a profile that you control
even when your contact address changes.

\begin{quote}
  \noindent
  \textbf{What's in a project?}

  When deciding what to store where,
  remember that your version control repository only stores part of what makes up your project.
  For example,
  while GitHub wikis are stored in ordinary git repositories, the issues and discussions live in GitHub's internal database;
  you can use the GitHub API to download their content,
  but (a) the result is intended for consumption by machines, not people,
  and (b) if you are doing this on short notice,
  the odds are high that other people are trying to do it at the same time.
  GitHub's servers may not be able to manage that load when you need them to,
  so use tools like \href{https://github.com/jlord/offline-issues}{offline-issues}
  to save hosted content as plain-text files regularly.
\end{quote}

Version control is not the only way to create and save copies of your work.
\href{https://www.softwareheritage.org/how-to-archive-reference-code/}{Software Heritage} archives software from multiple forges;
you can also snapshot the current state of repositories in a compressed archive file
(e.g., \texttt{.zip} or \texttt{.tar.gz})
and deposit those copies with the \href{https://osf.io/}{Open Science Foundation} (OSF),
\href{https://zenodo.org/}{Zenodo},
or \href{https://figshare.com/}{Figshare}.
(Zenodo even integrates with GitHub so that tagging a release on GitHub
automatically triggers deposit of a new archive on Zenodo,
but again,
this automation is only useful as long as both ends are accessible.)
Institutional,
\href{https://amt.coretrustseal.org/certificates/}{national},
or \href{https://safeguar.de/}{international} data repositories also enhance longevity.

Storing copies on someone else's computer is not the only option:
you can (and should) make copies of your projects on computers that you own.
Again,
this is a place where colleagues can help:
ask them to make copies on their computers in exchange for you making copies of theirs on yours.
Whatever you do,
document it \href{https://ropensci.org/blog/2022/03/22/safeguards-and-backups-for-github-organizations/}{as rOpenSci has}.

Similarly,
if you distribute your code as a package on \href{https://pypi.org/}{PyPI},
\href{https://anaconda.org/anaconda/conda}{Conda},
or \href{https://cran.r-project.org/}{CRAN},
that package can contain all of the source code.
Doing this also fosters community collaboration.
When setting these up, you should ideally ensure that control over the distribution channels for a package is not limited to a single developer.
However, this is not always possible.
For example,
while PyPI actively encourages this,
CRAN policies allow only single point of contact.

\section*{Rule 6: Encourage community adoption}

Publishing your code is not the same as publicizing it.
The more people who know about and rely on your project,
the greater the chances that someone will help keep it alive,
whether by working on it directly or supporting you in doing so.
If you have not named your project yet,
choose a name that does not hint at direct and exclusive affiliation to a single institution.
This can help if the code needs to be relocated,
and also makes contributors from other institutions feel welcome.
Additionally,
label entry-level issues
(e.g., to use the ``good first issue'' label on GitHub)
so that people can easily find a place to start \cite{Steinmacher2015}.

Going further,
there is a causal link between social media mentions and both new adopters and new contributors \cite{Fang2022}.
Announcing your project on mailing lists, forums, or social media,
and talking about it at conferences are therefore necessary but not sufficient:
since everyone else is doing this too,
it is hard to be heard above the noise.

There are many good guides to what you can do to promote your project \cite{Kuchner2011,BelliniSaibene2024}.
A short video showing how to use the software
or a slide deck that others can incorporate into their lectures
will increase uptake by lowering the cost of adoption.
Similarly,
a one-page website that opens with an elevator pitch
explaining who the software is for
and how it will improve their lives
is much more likely to lead to that crucial ``second glance''
than a list of publications.

Disseminating your code through tutorials at conferences is another effective strategy.
Tutorials are an opportunity for you,
as an author,
to describe your work in depth
and (more importantly) convince your audience to use it.
Networking through such events will enable you to build relationships with
the people most likely to care about your project,
and it is the best way to ensure that you land well after you leave your current position.

\begin{quote}
  \noindent
  \textbf{Bring people in}

  If you can, have at least one person outside your organization commit to your code.
  Such community contributions lead to joint ownership of intellectual property (IP),
  which makes it harder for any single institution to lock down your work.
  You can reinforce this by adding their name to the copyright statement in your license.
\end{quote}

Finally,
remember that it is not all about \emph{your} project.
If you ask others to help maintain your code,
be prepared to give them something in return:
a letter of reference,
help testing or documenting their project,
or co-authorship credit on the project and associated publications.

\section*{Rule 7: Do what you're supposed to (if you have time)}

Research software is increasingly recognized as a citable object \cite{Smith2016,Katz2021,Garijo2024},
and numerous books and research papers describe how to organize and run a research software project.
Rather than repeating what others have said
\cite{Sandve2013,Wilson2014},
this rule is a prioritized checklist of things to do
\emph{if you have time}.
If you do not,
please consult the next rule.

\begin{enumerate}

\item
  Create a DOI for each release of your software
  to ensure citability and retrievability even if the project repository becomes unavailable,
  and add a \href{https://citation-file-format.github.io/}{Citation.CFF}
  or \href{https://codemeta.github.io}{codemeta.json} file to your code \cite{Druskat2021}
  containing citation metadata.
  Tools such as \href{https://citation-file-format.github.io/cff-initializer-javascript/}{cffinit},
  \href{https://codemeta.github.io/codemeta-generator/}{codemeta generator},
  and \href{https://github.com/citation-file-format/citation-file-format/blob/main/README.md\#tools-to-work-with-citationcff-files-wrench}{others}
  can help you do this.

\item
  Submit your code for peer review and publication in a venue such as
  the \href{https://joss.theoj.org/}{Journal of Open Source Software},
  the \href{https://openresearchsoftware.metajnl.com/}{Journal of Open Research Software},
  the \href{http://www.jstatsoft.org}{Journal of Statistical Software},
  the \href{https://journal.r-project.org/}{R Journal},
  \href{https://ropensci.org/}{rOpenSci},
  or \href{https://www.pyopensci.org}{pyOpenSci}.

\item
  Document how to \emph{use} your software so that others can pick it up and use it if you are not available.
  In particular,
  add a \texttt{README.md} at the root of your repository
  that explains the purpose, setup, and use of your work,
  and add a tutorial showing how to use the major components of the code
  \cite{Lee2018b,Huybrechts2024,Littauer2025,Katz2025,Turing2025}.

\item
  Document any datasets your project depends on,
  and use CSV, JSON, Parquet, HDF5, or other widely-recognized formats rather than proprietary formats.
  Include the data in your project if possible;
  if that is not possible for legal reasons or because of the data's size,
  include URLs with data version IDs (or the date of the last download) in your documentation.

\item
  Describe your project's dependencies in a machine-readable way
  using Python's \href{https://packaging.python.org/en/latest/guides/writing-pyproject-toml/}{pyproject.toml} file,
  an npm \href{https://docs.npmjs.com/cli/v10/configuring-npm/package-json?v=true}{package.json} file,
  the \texttt{DESCRIPTION} and \texttt{NAMESPACE} files of an R package,
  a \href{https://openssf.org/technical-initiatives/sbom-tools/}{software bill of materials} (SBOM),
  or whatever else is standard for your language.

\item
  Write at least a few tests for your software
  so that people who want to use it (or contribute to it)
  can tell if it is working in their environment.
  These do not have to be formal unit tests \cite{Irving2021}:
  even a handful of ``this input should produce this output'' checks
  can save a lot of frustration \cite{Taschuk2017},
  Including sample inputs is also helpful,
  and removes dependence on external data sources which may themselves face continuity challenges.

\item
  Document how to \emph{contribute} to your software,
  e.g.,
  how to set up a development environment,
  run tests,
  and add new features.
  Such a guide leads to an increase in contributions \cite{Sholler2019},
  as does an explanation of the project's governance,
  i.e.,
  how decisions are made and who gets a voice in making them.

\end{enumerate}

Note that all of the suggestions above will also help with adoption,
for the same reason that a tidy restaurant attracts more customers than a messy one.
In addition,
if you follow the suggestions above when you have time,
there will be less for you to do when you don't
(as discussed in our next rule).

\section*{Rule 8: Do what you can (if you have to act quickly)}

``Good enough'' practices in research computing,
such as creating \texttt{.zip} archives of your project as an alternative to using version control,
may be more appropriate that ``best practices'' in an emergency \cite{Wilson2017}.
(By analogy,
a paramedic is not a doctor who cuts corners,
but rather someone who specializes in working under very different circumstances.)
If you are short on time:

\begin{enumerate}

\item
  Explanations of what the software \emph{does} and \emph{how to use it}
  take priority over documentation of its internals
  because (a) someone who knows the first two will be better able to figure out the third
  and (b) AI code assistants can do a reasonable job of helping newcomers navigate an unfamiliar code base
  if they are given enough context.

\item
  Similarly,
  a short screencast showing how you use your software
  is often faster to create than pages of documentation,
  and may actually be more useful if closed captions are attached
  to make the content searchable.
  Again,
  AI tools can help create those captions,
  though some editing is invariably required to clean up mistranslation of technical jargon.

\item
  Following on from Rule 5, you can \href{https://archive.softwareheritage.org/browse/search/}{search the Software Heritage Archive} with your repository url to check whether it has been saved.
  If your repository is not present, there is a \href{https://archive.softwareheritage.org/save/}{simple process} to submit it for archival.

\item
  If you do not have time to bring your project in line with best practices,
  try instead to clean it up by removing files that are no longer used,
  deleting sections of your setup instructions that no longer apply,
  closing bug reports that are no longer relevant,
  and so on.
  Eliminating clutter in this way will make what is useful easier to find.

\item
  Briefly document what happened.
  If you are the main developer working on a grant and you have moved on to something else,
  say that in the README.
  If the funding changed, then say that the work is unfunded.
  This information is often the most useful for users but the hardest to find.
  Saying it up front will help others understand how your work is positioned.

\item
  Following on from Rule~5,
  you can search the \href{https://archive.softwareheritage.org/browse/search/}{Software Heritage Archive}
  to check if your repository has been saved.
  If not,
  submitting it is a \href{https://archive.softwareheritage.org/save/}{simple process}.

\end{enumerate}

\section*{Rule 9: Talk about what you're doing}

All of the work that we have described is extra labor.
You will not want to do it,
particularly not when so many other things demand attention as well.
Make that work meaningful by helping others.
Tell your colleagues what you have learned about your institution's legal requirements,
about navigating its bureaucracy,
and about the least hurtful way to break the news to your students, collaborators, and users.
Share your own rules on social media
(preferably Mastodon rather than X or Bluesky,
as the latter two are single points of institutional failure),
and share things that \emph{haven't} worked as well
so that other people can avoid the same pitfalls.

Sharing your experience during a difficult time is valuable even if your only audience is yourself.
Writing down what happened, what you did, how you felt, what you have learned, and what you hope for
are evidence-based strategies for protecting your mental health in stressful times
and for healing afterwards \cite{Pennebaker2016,Cullen2022}.

Most importantly,
take the time to grieve.
Having years of work come to an abrupt end will take a toll on your mental health;
many of your colleagues will be suffering as well,
and those who haven't been hit (yet) may be affected by survivors' guilt.
Be angry,
be sad,
but remember that in the aftermath of a disaster---natural or man-made---people
often become altruistic, resourceful, and brave \cite{Yang2024}.

\section*{Rule 10: Organize}

There is little point surviving today's flood if you drown in tomorrow's.
The best strategy is to take preventive action,
and the best way to do that in a research context
is to become active in professional associations
and push them to take meaningful action.

Many readers of this paper will find the thought of doing this uncomfortable,
in large part because 
voluntary organizations have seen a slow but steady decline in membership over several decades \cite{Putnam2020}.
Although there are many contributing factors,
the net result is that professionals aged 30--60
are less likely to be involved in civil society and the political process than the generation before them.
This disengagement effectively hands power to extremists and special interest groups
and enables them to seize control of civil institutions in order to advance their agendas \cite{BuenoDeMesquita2022}.
To see the impact,
ask yourself who is more likely to run for a seat on your local school board:
someone with a doctorate in epidemiology or an anti-vax conspiracy theorist?

We have written this paper because we believe that scientific inquiry is a public good and worth fighting for.
The present climate requires us to focus on preserving what we can,
but playing defense is just a way to lose more slowly.
If we work together and accept that we and our institutions need to change,
we can create a world in which research does not just survive,
but thrives.

\bibliography{code-rules}

\begin{thebibliography}{10}

\bibitem{Pearson2025}
Pearson H, Ledford H, Hutson M, Van~Noorden R.
\newblock Exclusive: the most-cited papers of the twenty-first century.
\newblock Nature. 2025;640(8059):588--592.
\newblock doi:{10.1038/d41586-025-01125-9}.

\bibitem{Carver2022}
Carver JC, Weber N, Ram K, Gesing S, Katz DS.
\newblock A survey of the state of the practice for research software in the
  United States.
\newblock PeerJ Computer Science. 2022;8:e963.
\newblock doi:{10.7717/peerj-cs.963}.

\bibitem{Woodward2025}
Woodward A, Leeder S.
\newblock Making science great again. Or not.
\newblock International Journal of Epidemiology. 2025;54(2).
\newblock doi:{10.1093/ije/dyaf029}.

\bibitem{Nature2019}
{Nature}.
\newblock Mexican science suffers under debilitating budget cuts.
\newblock Nature. 2019;572:294--295.
\newblock doi:{10.1038/d41586-019-02332-x}.

\bibitem{Nature2021}
{Nature}.
\newblock Brazil’s scientists face 90
\newblock Nature. 2021;598:566.
\newblock doi:{10.1038/d41586-021-02882-z}.

\bibitem{Nature2023}
{Nature}.
\newblock Science is under threat in Argentina — we must call out the danger.
\newblock Nature. 2023;622:433.
\newblock doi:{10.1038/d41586-023-03228-7}.

\bibitem{RatRisk2024}
Inspireurope+. Researchers at Risk: An Update on National level Actions in
  Europe 2024; 2024.
\newblock Available from:
  \url{https://sareurope.eu/inspireurope/inspireurope-publications-policy/}.

\bibitem{Nature2025}
{Nature}.
\newblock Trump 2.0: An assault on science anywhere is an assault on science
  everywhere.
\newblock Nature. 2025;639(8053):7--8.
\newblock doi:{10.1038/d41586-025-00562-w}.

\bibitem{Thakur2025}
Malviya-Thakur A, Milewicz R, Jahanshahi M, Paganini L, Vasilescu B, Mockus A.
\newblock Scientific open-source software is less likely to become abandoned
  than one might think! Lessons from curating a catalog of maintained
  scientific software.
\newblock In: ACM International Conference on the Foundations of Software
  Engineering (FSE); 2025.Available from:
  \url{https://arxiv.org/abs/2504.18971}.

\bibitem{Perkel2023}
Perkel JM.
\newblock How to make your scientific data accessible, discoverable and useful.
\newblock Nature. 2023;618(7967):1098--1099.
\newblock doi:{10.1038/d41586-023-01929-7}.

\bibitem{Nikiforuk2013}
Nikiforuk A. Harper government shuts down 'world class' collection on
  freshwater science and protection;.
\newblock Available from:
  \url{https://thetyee.ca/News/2013/12/09/Dismantling-Fishery-Library/}.

\bibitem{DelBoca2024}
{Del Boca} P. Disorderly Transition: Preserving the Argentine Government's Open
  Data; 2024.
\newblock presented at {csv,conf} 8.
\newblock Available from: \url{https://www.youtube.com/watch?v=8dUyi4OYAdM}.

\bibitem{Katz2018}
Katz DS, Niemeyer KE, Smith AM.
\newblock Publish your software: Introducing the Journal of Open Source
  Software (JOSS).
\newblock Computing in Science \& Engineering. 2018;20(3):84--88.
\newblock doi:{10.1109/MCSE.2018.03221930}.

\bibitem{Ham2019}
Ham D, Hargreaves JC, Kerkweg A, Roche DM, Sander R.
\newblock The publication of geoscientific model developments v1.2.
\newblock Geosci Model Dev. 2019;12(6).
\newblock doi:{10.5194/gmd-12-2215-2019}.

\bibitem{Scott1987}
Scott JC.
\newblock Weapons of the Weak: Everyday Forms of Peasant Resistance.
\newblock Yale University Press; 1987.

\bibitem{Torr2005}
Torr P.
\newblock Demystifying the threat modeling process.
\newblock IEEE Security \& Privacy. 2005;3(5):66--70.
\newblock doi:{10.1109/MSP.2005.119}.

\bibitem{Barker2010}
Barker K.
\newblock At the Helm: Leading Your Laboratory.
\newblock Cold Spring Harbor Laboratory Press; 2010.

\bibitem{Cohen2018}
Cohen CM, Cohen SL.
\newblock Lab Dynamics: Management and Leadership Skills for Scientists.
\newblock 3rd ed. Cold Spring Harbor Laboratory Press; 2018.

\bibitem{deRSE25}
Yehudi Y, Cashman M, Felderer M, Goedicke M, Hasselbring W, Katz DS, et~al.
\newblock Towards Defining Lifecycles and Categories of Research Software.
\newblock deRSE25 Conference Proceedings. 2025;doi:{10.5281/zenodo.15002661}.

\bibitem{Morin2012}
Morin A, Urban J, Sliz P.
\newblock A Quick Guide to Software Licensing for the Scientist-Programmer.
\newblock PLOS Computational Biology. 2012;8(7):e1002598.
\newblock doi:{10.1371/journal.pcbi.1002598}.

\bibitem{Fogel2020}
Fogel K. Producing Open Source Software (updated); 2020.
\newblock Available from: \url{https://producingoss.com/}.

\bibitem{Fortunato2021}
Fortunato L, Galassi M.
\newblock The case for free and open source software in research and
  scholarship.
\newblock Phil Trans Royal Soc A. 2021;379(2197).
\newblock doi:{10.1098/rsta.2020.0079}.

\bibitem{Gomes2022}
Gomes DGE, Pottier P, Crystal-Ornelas R, Hudgins EJ, Foroughirad V,
  Sánchez-Reyes LL, et~al.
\newblock Why don’t we share data and code? Perceived barriers and benefits
  to public archiving practices.
\newblock Proceedings of the Royal Society B: Biological Sciences.
  2022;289(1987).
\newblock doi:{10.1098/rspb.2022.1113}.

\bibitem{Langille2010}
Langille MGI, Eisen JA.
\newblock BioTorrents: A File Sharing Service for Scientific Data.
\newblock PLoS ONE. 2010;5(4):e10071.
\newblock doi:{10.1371/journal.pone.0010071}.

\bibitem{Tamburri2020}
Tamburri DA, Blincoe K, Palomba F, Kazman R.
\newblock {``The Canary in the Coal Mine…''} A cautionary tale from the
  decline of SourceForge.
\newblock Software: Practice and Experience. 2020;50(10):1930--1951.
\newblock doi:{10.1002/spe.2874}.

\bibitem{Steinmacher2015}
Steinmacher I, Conte T, Gerosa MA, Redmiles D.
\newblock Social Barriers Faced by Newcomers Placing Their First Contribution
  in Open Source Software Projects.
\newblock In: Proc. {CSCW'15}. {ACM}; 2015.

\bibitem{Fang2022}
Fang H, Lamba H, Herbsleb J, Vasilescu B.
\newblock ``This is damn slick!'': estimating the impact of tweets on open
  source project popularity and new contributors.
\newblock In: Proc.\ 44th International Conference on Software Engineering;
  2022. p. 2116--29.

\bibitem{Kuchner2011}
Kuchner MJ.
\newblock Marketing for Scientists: How to Shine in Tough Times.
\newblock Island Press; 2011.

\bibitem{BelliniSaibene2024}
{Bellini Saibene} Y, Salmon M. Marketing Ideas For Your Package; 2024.
\newblock Available from:
  \url{https://ropensci.org/blog/2024/03/07/package-marketing/}.

\bibitem{Smith2016}
Smith AM, Katz DS, Niemeyer KE, Group FSCW.
\newblock Software citation principles.
\newblock PeerJ Computer Science. 2016;2:e86.
\newblock doi:{10.7717/peerj-cs.86}.

\bibitem{Katz2021}
Katz DS, Chue~Hong NP, Clark T, et~al.
\newblock Recognizing the value of software: a software citation guide [version
  2; peer review: 2 approved].
\newblock F1000Research. 2021;9:1257.
\newblock doi:{10.12688/f1000research.26932.2}.

\bibitem{Garijo2024}
Garijo D, Arroyo M, Gonzalez E, Treude C, Tarocco N.
\newblock Bidirectional paper-repository tracing in software engineering.
\newblock In: Proceedings of the 21st International Conference on Mining
  Software Repositories; 2024. p. 642--646.

\bibitem{Sandve2013}
Sandve GK, Nekrutenko A, Taylor J, Hovig E.
\newblock Ten Simple Rules for Reproducible Computational Research.
\newblock PLOS Computational Biology. 2013;9(10):e1003285.
\newblock doi:{10.1371/journal.pcbi.1003285}.

\bibitem{Wilson2014}
Wilson G, Aruliah DA, Brown CT, Chue~Hong NP, Davis M, Guy RT, et~al.
\newblock Best Practices for Scientific Computing.
\newblock PLOS Biology. 2014;12(1):e1001745.
\newblock doi:{10.1371/journal.pbio.1001745}.

\bibitem{Druskat2021}
Druskat S, Spaaks JH, Chue~Hong N, Haines R, Baker J, Bliven S, et~al..
  Citation File Format; 2021.

\bibitem{Lee2018b}
Lee BD.
\newblock Ten simple rules for documenting scientific software.
\newblock PLOS Computational Biology. 2018;14(12):e1006561.
\newblock doi:{10.1371/journal.pcbi.1006561}.

\bibitem{Huybrechts2024}
Huybrechts P, Trekels M, Abraham L, Desmet P. B-Cubed software development
  guide; 2024.
\newblock Available from:
  \url{https://docs.b-cubed.eu/guides/software-development/}.

\bibitem{Littauer2025}
Littauer R. Standard Readme; 2025.
\newblock Available from: \url{https://github.com/RichardLitt/standard-readme}.

\bibitem{Katz2025}
Katz DS, Forbes M, Silen L, Curcuru S, Hucka M, Tang Y, et~al.. Open-source
  software project documents; 2025.
\newblock Available from: \url{https://github.com/corsa-center/oss-documents}.

\bibitem{Turing2025}
{The Turing Way Community}. The Turing Way: A handbook for reproducible,
  ethical and collaborative research; 2025.

\bibitem{Irving2021}
Irving D, Hertweck K, Johnston L, Ostblom J, Wickham C, Wilson G.
\newblock Research Software Engineering with Python: Building Software that
  Makes Research Possible.
\newblock CRC Press/Taylor and Francis; 2021.

\bibitem{Taschuk2017}
Taschuk M, Wilson G.
\newblock Ten simple rules for making research software more robust.
\newblock PLOS Computational Biology. 2017;13(4):e1005412.
\newblock doi:{10.1371/journal.pcbi.1005412}.

\bibitem{Sholler2019}
Sholler D, Steinmacher I, Ford D, Averick M, Hoye M, Wilson G.
\newblock Ten simple rules for helping newcomers become contributors to open
  projects.
\newblock {PLOS} Computational Biology. 2019;15(9):e1007296.
\newblock doi:{10.1371/journal.pcbi.1007296}.

\bibitem{Wilson2017}
Wilson G, Bryan J, Cranston K, Kitzes J, Nederbragt L, Teal TK.
\newblock Good Enough Practices in Scientific Computing.
\newblock {PLOS} Computational Biology. 2017;13(6):1--20.
\newblock doi:{10.1371/journal.pcbi.1005510}.

\bibitem{Pennebaker2016}
Pennebaker JW, Smyth JM.
\newblock Opening Up by Writing It Down: How Expressive Writing Improves Health
  and Eases Emotional Pain.
\newblock 3rd ed. Guilford Publications; 2016.

\bibitem{Cullen2022}
Cullen L, Hanrahan K, Farrington M, Tucker S, Edmonds S.
\newblock Evidence-Based Practice in Action: Comprehensive Strategies, Tools,
  and Tips From University of Iowa Hospitals \& Clinics.
\newblock 2nd ed. Sigma Theta Tau International; 2022.

\bibitem{Yang2024}
Yang Y.
\newblock The Effect of Disasters on Altruistic Behavior and its Mechanisms.
\newblock Journal of Education, Humanities and Social Sciences.
  2024;26:572–577.
\newblock doi:{10.54097/xng2q069}.

\bibitem{Putnam2020}
Putnam RD.
\newblock Bowling Alone: The Collapse and Revival of American Community.
\newblock 2nd ed. Simon \& Schuster; 2020.

\bibitem{BuenoDeMesquita2022}
{Bueno de Mesquita} B, Smith A.
\newblock The Dictator's Handbook: Why Bad Behavior is Almost Always Good
  Politics.
\newblock PublicAffairs; 2022.

\end{thebibliography}

\end{document}